\documentclass{llncs}
\usepackage{algorithm}
\usepackage{algpseudocode}
\usepackage{enumitem}
\usepackage{multirow,array}
\usepackage{graphicx}
\usepackage{xcolor}
\usepackage{tikz}
\usepackage{rotating} 
\usepackage{comment}
\usepackage{subfigure}
\graphicspath{{./pics/}}

\usepackage{listings}

\definecolor{cmtgray}{gray}{0.45}
\definecolor{cmtmgray}{gray}{0.20}
\lstdefinestyle{customc}{
  belowcaptionskip=.3\baselineskip,
  escapeinside={/*@}{@*/},
  breaklines=true,
  xleftmargin=\parindent,
  language=C++,
  showstringspaces=false,
  breakatwhitespace=true,
  basicstyle=\ttfamily\small,
  keywordstyle=\bfseries\color{green!40!black}\textbf,
  commentstyle=\color{cmtgray}\ttfamily,
  identifierstyle=\color{black!90!white},
  stringstyle=\color{purple},
  captionpos=b,
 morekeywords={},
}
\newcommand{\prg}[1]{\texttt{#1}}
\newcommand{\prgop}[1]{\textcolor{cyan}{\texttt{#1}}}
\newcommand{\prgvar}[1]{\textcolor{blue}{\texttt{#1}}}
\newcommand{\prgvarx}[1]{\textcolor{red}{\texttt{#1}}}

\begin{document}

\title{TaskUniVerse: A Task-Based Unified Interface for Versatile Parallel Execution}
\titlerunning{TaskUniVerse}

\author{Afshin Zafari}
\institute{Uppsala  University,\\Department of Information Technology,\\Division of Scientific Computing\\
L{\"a}gerhyddsv{\"a}gen 2, 752 37 Uppsala, Sweden\\
\email{afshin.zafari@it.uu.se}
}


\maketitle
\begin{abstract}
Task based parallel programming has shown competitive outcomes in many aspects of parallel programming such as efficiency, performance, productivity and scalability. Different approaches are used by different software development frameworks to provide these outcomes to the programmer, while making the underlying hardware architecture transparent to her. However, since programs are not portable between these frameworks, using one framework or the other is still a vital decision by the programmer whose concerns are expandability, adaptivity, maintainability and interoperability of the programs.
In this work, we propose a unified programming interface that a programmer can use for working with different task based parallel frameworks transparently. In this approach we abstract the common concepts of task based parallel programming and provide them to the programmer in a single programming interface uniformly for all frameworks.
We have tested the interface by running programs which implement matrix operations within frameworks that are optimized for shared and distributed memory architectures and accelerators, while the cooperation between frameworks is configured externally with no need to modify the programs. Further possible extensions of the interface and future potential research are also described.
\end{abstract}


\section{Introduction}
Task based parallel programming has experienced a great increase in the past decade due to its competitive outcomes in performance and productivity. The key to success for task based approaches is the abstract view of a program as a set of operations and data. This abstraction allows for programs to be written sequentially using tasks and data whenever an operation is to be performed on program variables/identifiers. When such a program runs, the tasks corresponding to the operations in the program are submitted to the background task-based framework run-time system where they are scheduled for parallel execution. This separation of a written program and its underlying tasks, enables the providers of the task-based frameworks to use different approaches for finding the optimal solution to the scheduling problem of running the tasks on the available computing resources. 

Different techniques are provided by task based programming frameworks to the programmer for writing programs. The StarPU \cite{starpu:main} and OmpSs \cite{ompss:main} frameworks extend the C compiler and allow the programmer to use compiler directives to define C functions as tasks kernels and describe their data dependencies. The PaRSEC \cite{parsec:main,parsec:ptg} framework provides tools and utilities to analyze a program written in a special language that describes tasks and data dependencies and us	es a source to source compiler to translate the optimal solution into a C code for compilation. The DuctTeip \cite{dt:main}, Chunks and Tasks \cite{cht:main} and also StarPU \cite{starpu:mpi} frameworks provide Application Programming Interface (API) for defining data and tasks to run in a distributed memory environment. The SuperGlue framework \cite{sg:main} provides a headers only C++ library for creating tasks and running them on multi-core processors.

These frameworks have individually shown very good results in terms of performance, scalability and  productivity and in a wide spectrum of scientific applications such as solving partial differential equations (PDE) \cite{sg:sw}, N-Body problems using Fast Multipole Method (FMM)  \cite{spu:fmm,sg:fmm,sg:turin}, simulating stochastic discrete events \cite{sg:stochastic,sg:stochastic2}, Conjugate Gradient method \cite{spu:cg}, Finite Element Method (FEM) applications \cite{spu:fem}, chemistry applications \cite{parsec:chemi}, seismic  applications \cite{spu:seismic} and image processing \cite{parsec:imag}. They have also shown the feasibility and benefits of using task based approaches for sparse data structures, \cite{starpu:sparse,cht:app}.

There are also attempts to join pairs of the task based frameworks to combine benefits of both. The StarPU and PaRSEC frameworks joined to provide task parallelism in clusters of heterogeneous processors \cite{spu:magma,starpu:sparse}. The DuctTeip and SuperGlue frameworks joined \cite{dt:main} to implement hierarchical task submission and execution in hybrid distributed and shared memory environments. 

While these achievements seem promising for using task based programming models, the choice of a proper task based framework may still be risky when impacts on legacy software due to software and hardware changes may occur due to differences in framework implementations.

In this paper, we address these issues by proposing a unified task based programming (UTP) model in which any number of task based frameworks can be used to run a single application in many parallel environments. This section continues with explaining the motivation for designing such a programming model. The overview, implementation and programming of the UTP model are described in Sections \ref{sec:overview}--\ref{sec:programming}.
Section \ref{sec:experiments} shows the performance results of executing a Cholesky factorization program in different parallel computing environments. The last section is devoted to discussion and conclusions.

\subsection{Motivation}\label{sec:motiv}
The number of applications that use task based programming approaches is increasing and more attempts to join two or more task based frameworks to exploit different advantages can be foreseen. These achievements for the task based parallel programming approaches encourage the application domain scientists to see that the solution of the problems they are trying to solve can be implemented in an efficient and scalable way on thousands of processors. However, choosing frameworks for implementing  solutions for a specific application domain is still a vital decision for the expert end users in that domain. 

The basic factors influencing the choice of framework(s) are richness and flexibility. When needs of a scientific application span a wide spectrum of software and hardware varieties, it becomes hard or impossible to find a single framework to address them. Also the investment of developing an application even on top of a mixture of frameworks has to be secured against probable risks of future changes in underlying software and hardware. In this paper we show how these issues can be addressed by the UTP model through a unified task programming interface.

\section {The UTP Model}
\subsection{Overview of the UTP Model}\label{sec:overview}
The UTP programming model is designed to provide an abstraction for common structures and behaviors of the task based frameworks mentioned above. In this abstract view, all the operations performed by a program on its data are replaced by tasks and special data types (e.g., handles or descriptors) representing the program data. Instead of running the operations immediately, the corresponding tasks are submitted to the frameworks' runtime where their dependencies are tracked and ready tasks are executed in parallel. Actual kernel computations of a ready task are performed through call back mechanisms which are introduced to the frameworks at task creation.

In order to have a single interface for cooperation, the UTP model requires the frameworks to implement predefined interfaces for data definition and task creation, submission, execution and completion. These interfaces unify the cooperation of UTP with any other compliant framework via conversations of \emph{generic} data and tasks regardless of the concrete instances inside the framework. A central \emph{dispatcher} in the UTP model orchestrates the flow of tasks and data between the program and frameworks by connecting the interfaces of one framework to another. The dispatcher submits tasks to the frameworks and they will notify the dispatcher when the tasks are ready to execute or finished. 

The UTP model divides the software stack into three layers, as shown in Fig.~\ref{fig:overview},  where the UTP interface in the middle decouples the application layer at the top from the task based frameworks at the bottom which in turn hide the technical details of parallel programming for the underlying hardware. In the UTP model, the taskified versions of the operations are provided by the middle layer to the application layer via ordinary subroutines while on the other side of the middle layer, the generic tasks move back and forth to the frameworks or their UTP compliant wrappers. Therefore, the UTP at  middle layer translates program operations to tasks and data and distributes them properly to available  task based frameworks through a generic interface.

Different task flows between dispatcher and task based frameworks can be \emph{configured} in the UTP model by specifying which two interfaces of frameworks are to be interconnected via the dispatcher. For example, the \emph{task-execution} interface of one framework can be connected to the \emph{task-creation} (submission) interface of another to enable hierarchical task management in which, when tasks get ready to execute at higher levels, they split into child tasks and are submitted to the framework at the next level of the hierarchy. This configurable task flow graph between the \emph{program} and the \emph{wrappers} allows a single program to run in different parallel computing environments by different task based frameworks.

\subsection{Implementation of the UTP Model}\label{sec:implementation}
The UTP model provides the necessary data structures and interfaces to the application layer for defining data and performing operations on them. These interfaces also enable \emph{partitioning} the data into parts and \emph{splitting} an operation into child tasks. Usages of data definition and partitioning interfaces by the program are propagated to all the task based frameworks to manage their own internal data types.

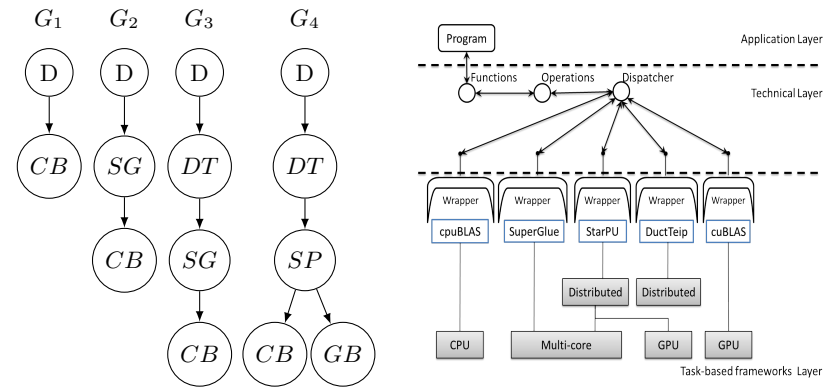
\begin{figure}
\subfigure[Possible task flow graphs $G_1$--\,$G_4$. $D$ is the \emph{dispatcher} and $DT$, $SG$, $SP$ are wrappers around the DuctTeip, SuperGlue and StarPU frameworks, respectively;  $CB$ and $GB$ are wrappers around BLAS libraries on CPU and GPU, respectively.]{
\label{fig:topo}
\begin{tikzpicture}[xscale=0.5]
    \node[shape=circle,draw=black] (D) at (1,0) {D};
    \node[shape=circle,draw=black] (S1) at (1 ,-1.25) {$CB$};

    \node[shape=circle,draw=none]  at (1 ,.71) {$G_1$};
    \path [-latex] (D) edge   (S1);

    \node[shape=circle,draw=black] (aD) at (3,0) {D};
    \node[shape=circle,draw=black] (aS1) at (3 ,-1.25) {$SG$};
    \node[shape=circle,draw=black] (aS2) at (3 ,-2.5) {$CB$};

    \node[shape=circle,draw=none]  at (3 ,.71) {$G_2$};
    \path [-latex] (aD) edge   (aS1);
    \path [-latex] (aS1) edge   (aS2);

    \node[shape=circle,draw=black] (bD) at (5,0) {D};
    \node[shape=circle,draw=black] (bS1) at (5 ,-1.25) {$DT$};
    \node[shape=circle,draw=black] (bS2) at (5 ,-2.5) {$SG$};
    \node[shape=circle,draw=black] (bS3) at (5 ,-3.75) {$CB$};

    \node[shape=circle,draw=none]  at (5 ,.71) {$G_3$};

    \path [-latex] (bD) edge   (bS1);
    \path [-latex] (bS1) edge   (bS2);
    \path [-latex] (bS2) edge   (bS3);

    \node[shape=circle,draw=black]  (dD) at (7.8,0) {D};
    \node[shape=circle,draw=black] (dS1) at (7.8,-1.25) {$DT$};
    \node[shape=circle,draw=black] (dS2) at (7.8,-2.5) {$SP$};
    \node[shape=circle,draw=black] (dS3) at (7.0,-3.75) {$CB$};
    \node[shape=circle,draw=black] (dS4) at (8.81,-3.75) {$GB$};

    \node[shape=circle,draw=none]  at (7.8,.71) {$G_4$};
    \path [-latex] (dD) edge   (dS1);
    \path [-latex] (dS1) edge   (dS2);
    \path [-latex] (dS2) edge   (dS3);
    \path [-latex] (dS2) edge   (dS4);

\end{tikzpicture}
}\quad\subfigure[The overview of the UTP model. The UTP interface with task-based frameworks unifies the cooperation of Dispatcher and any framework run-time. A single program in the Application Layer can be executed in various parallel environments using combinations of frameworks (or their wrappers).]{
\includegraphics[width=.45\textwidth,height=.25\textheight]{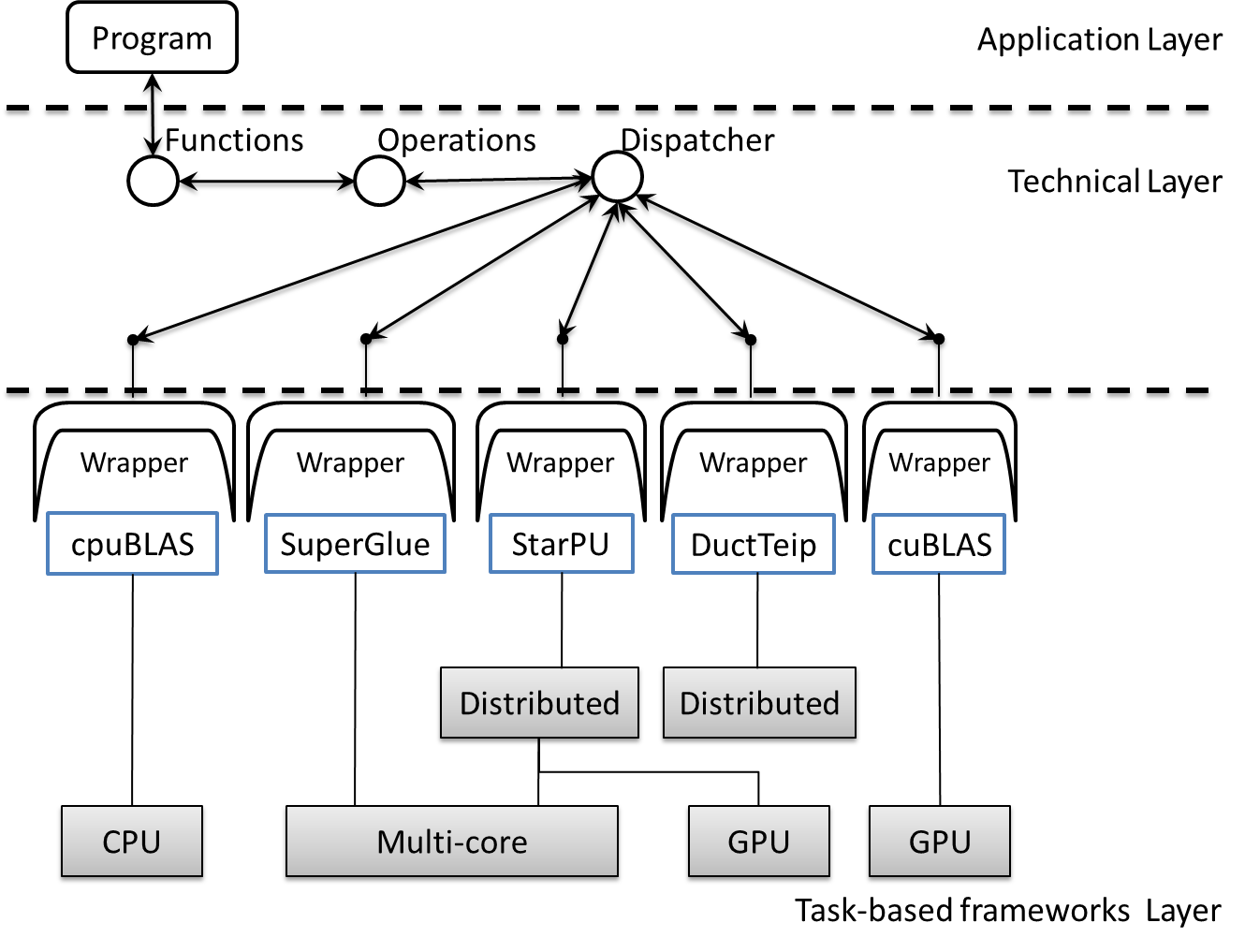}
\label{fig:overview}
}
\caption{Cooperation between the UTP and the task based frameworks.}
\end{figure}

For running a program using various frameworks, the UTP model also contains implemented wrappers around the SuperGlue, StarPU and DuctTeip task based frameworks, Fig.~\ref{fig:overview}. There are also cpuBLAS and cuBLAS wrappers around Basic Linear Algebra Subprograms (BLAS) and Linear Algebra PACKage (LAPACK) libraries for CPU and GPU devices, respectively, that can be used in the task flow graph for running the actual computations of tasks on the corresponding devices. The tasks submitted to these two wrappers are immediately executed and their completions are reported back to the dispatcher. 

Figure~\ref{fig:topo} shows some examples of possible and practical configurations of task flow graphs $G_1$--\,$G_4$ for running a program whose operations on data are totally decomposable into BLAS subroutines. 
The edge from the program to the dispatcher $D$ is common for all the graphs and is not shown in the figure.
The flow of tasks and data between the dispatcher $D$ and the wrappers can be configured to determine the hierarchy of data and tasks and the corresponding responsible frameworks.

The nodes in the graphs shown in Fig.~\ref{fig:topo} are the dispatcher $D$ and the wrappers around the task based frameworks. A directed edge from the dispatcher $D$ to node $w$ denotes that the tasks coming from the program to the dispatcher are forwarded to wrapper $w$. The edges between wrappers $w_1$ and $w_2$ show that ready tasks at $w_1$ are split into sub tasks by the dispatcher and are submitted to $w_2$, and task completions at $w_2$ are reported back to $w_1$ via the dispatcher. 

Using configurations $G_1$--\,$G_4$ the program can run in distributed-/shared- memory and heterogeneous (with accelerators) computing environments. Using the $G_1$ configuration, it can run sequentially on one CPU since tasks submitted to the dispatcher are forwarded to and immediately executed by the  cpuBLAS wrapper. In the $G_2$ configuration, the same program can run on multi-core systems using the SuperGlue wrapper for managing submitted tasks on available cores and using the BLAS library wrapper for running the tasks on individual cores.  The $G_3$ configuration is constructed by adding the DuctTeip wrapper on top of several wrappers and enables the program to run in a cluster of computing nodes where DuctTeip is managing data and tasks in distributed memory environments. In the $G_4$ configuration, the program can run in a cluster of computing nodes with heterogeneous CPU/GPU processors using StarPU for managing tasks on both CPU and GPU. 

\subsection{Programming in the UTP Model}\label{sec:programming}
In the three layers view of the UTP model, the required programming at the bottom layer is already done in the UTP framework and provided as wrappers around some existing and frequently used task based frameworks. Further development of wrappers can be performed by framework providers or users by implementing the unified interface.

The programming at the middle layer, consists of implementing the translation and splitting of operations into tasks or sub tasks. This can be done by implementing predefined interfaces (e.g., the \prg{split} method of an \prgop{operation} object) which will be used by the dispatcher during the program execution. User friendly functions can hide the technical details of the task and operations from the the programmer at the application layer. The programming at the application layer consists of defining data and their partitions and calling functions provided by the technical layer to manipulate the data. 

The programming in the UTP model is exemplified by implementing a block Cholesky matrix factorization called POTRF (POsitive definite matrix TRiangular Factorization) in BLAS/LAPACK terminology. The program at the application layer (shown in Fig.~\ref{prg:main}, lines \ref{line:mainstart}--\ref{line:mainend}) defines the input/output matrix \prgvarx{A} and its partitioning in two subsequent hierarchical levels (\prgvar{b1} and \prgvar{b2}) with parameters read from the command line and passes it to the \texttt{utp\_cholesky} function which is implemented in the technical layer (Fig.~\ref{prg:main}, lines \ref{line:cholstart}--\ref{line:cholend}).

\begin{figure}
\subfigure[The main program in the UTP model for implementing a Cholesky factorization of input matrix \prgvarx{A} (lines \ref{line:mainstart}--\ref{line:mainend}), the \prg{utp\_cholesky} function provided by the UTP technical layer (lines \ref{line:cholstart}--\ref{line:cholend}), and the run method of the \prgop{upotrfo} operation (lines \ref{line:blasrunstart}--\ref{line:blasrunend}). ]{
\begin{lstlisting}[mathescape,numbers=left,boxpos=b]
// unified_cholesky.cpp /*@\label{line:mainstart}@*/
#include "utp.hpp"
int main(int argc, char **argv){
  // UTP start
  utp_initialize(argc,argv);   

  int N, b1, b2;
  // Get dimensions and partitions
  // from command line
  utp_get_parameters(N,b1,b2);

  GData A(N, N, b1, b2);
  utp_cholesky(A);                 

  // UTP waits for all tasks finished
  utp_finalize();             
}/*@\label{line:mainend}@*/
/*-----------------------------*/
//utp_chol.cpp /*@\label{line:cholstart}@*/
#include "utp.hpp"
void utp_cholesky(GData &A){
  POTRFTask *potrf = new /*@\label{line:potrftask}@*/
    POTRFTask(upotrfo,NULL,A);
  dispatcher->submit_task(potrf);
}/*@\label{line:cholend}@*/
/*-----------------------------*/
// cpuBLAS_wrapper.cpp /*@\label{line:blasrunstart}@*/
#include "utp.hpp"
...
upotrfo::run(GTask *t){
  GData &A = *t->args[0];
  double *mem = A.get_memory();/*@\label{line:mem}@*/
  int info, N = A.get_rows_count();/*@\label{line:lldsz}@*/
    
  dpotrf('L',N,mem,N,&info);
  dispatcher->task_finished(t);
}/*@\label{line:blasrunend}@*/
\end{lstlisting}
\label{prg:main}
}\quad\subfigure[The splitting method of the POTRF operation object in the UTP programming model where child tasks (SYRK, GEMM, POTRF and TRSM) for the parent task \prgvar{p} are created and submitted to the \prgvar{dispatcher} with their corresponding \prgop{u<name>o} operation objects.]{
\begin{lstlisting}[mathescape,boxpos=b]
void upotrfo::split(GTask *p){  
  //unpack arguments of t to A
  GData &A = p->args[0];
  
  int n = A.row_part_num();  
  for(int i = 0; i<n; i++){
    for(int j = 0; j<i; j++){
      // submit task for $ \color{cmtmgray}A_{ii}=A_{ij}A_{ij}^T$
      SYRKTask *syrk = new
	SYRKTask(usyrko,p,A(i,j),A(i,i));
      dispatcher->submit_task(syrk);
      for(int k = i+1; k<n; k++){
	// submit task for $ \color{cmtmgray}A_{ki}=A_{ki}+A_{kj}A_{ij}$
	GEMMTask *gemm = new
	  GEMMTask(ugemmo,p,
		   A(k,j),A(i,j),A(k,i));
	dispatcher->submit_task(gemm);
      }
    }
    // submit task for $ \color{cmtmgray}A_{ii}\rightarrow LL^T$
    POTRFTask *potrf = new
      POTRFTask(upotrfo,p,A(i,i));
    dispatcher->submit_task(potrf);
    for(int j = i+1; j<n; j++){
      // submit task for $ \color{cmtmgray}A_{ji}=A_{ii}^{-1}A_{ji}$
      TRSMTask *trsm = new
	TRSMTask(utrsmo,p,A(i,i),A(j,i));
      dispatcher->submit_task(trsm);
    }
  }
}/*@\label{line:taskified_end}@*/
\end{lstlisting}
\label{prg:upotrfo}
}
\caption{Source codes of the Cholesky factorization in the UTP model.}
\end{figure}

The \prg{<name>Task} objects in Figs. \ref{prg:main} and \ref{prg:upotrfo} are subclasses of a generic task class in the UTP model whose constructors accept an \prgop{Operation} object, a pointer to the parent task and data arguments of the task. The created \prg{POTRFTask} in the \prg{utp\_cholesky} function (Fig.~\ref{prg:main}, line \ref{line:potrftask})
corresponds to the operation object \prgop{upotrfo} with no parent task.

The \prgop{Operation} objects in the UTP model are responsible for splitting an operation into child tasks that manipulate the partitions of the parent task's data arguments. Figure~\ref{prg:upotrfo} shows  the \prg{split} method of the \prgop{upotrfo} operation which in nested loops manipulates the partitions of input argument \prgvarx{A} using the indexing interface \prgvarx{A}\prg{(r,c)} to access the partition at row \prg{r} and column \prg{c} of \prgvarx{A}.

At the lowest level of the task hierarchy, when a task is submitted by the dispatcher to \prg{<cpu/cu>BLAS} wrappers the \prg{run} method of the task's operation is invoked where the BLAS/LAPACK routine is immediately called and task completion is reported back to the dispatcher, as shown in Fig.~\ref{prg:main} lines \ref{line:blasrunstart}--\ref{line:blasrunend}.

\section{Experiments}\label{sec:experiments}
To demonstrate the productivity gained by the UTP programming model, the Cholesky matrix factorization algorithm is implemented and executed with different matrix sizes on different parallel computing resources. In all these experiments the program at the application layer is written once and executed in different configurations for different underlying parallel hardware.

These programs were executed in 
the HPC2N computer cluster Kebnekaise using 32 nodes, each with two Intel Xeon E5-2690v4 CPU with 14 cores and with two NVIDIA K80 with 4992 cores. 
 The programs are all written in C++, compiled with Intel compiler 17.0.1 and Intel MPI version 2017 Update 1 and use Intel MKL for BLAS/LAPACK routines. 

\begin{figure}
\subfigure[Executing the Cholesky factorization in one computing node with or without GPU.]
{
\includegraphics[width=.5\textwidth,height =.25\textheight]{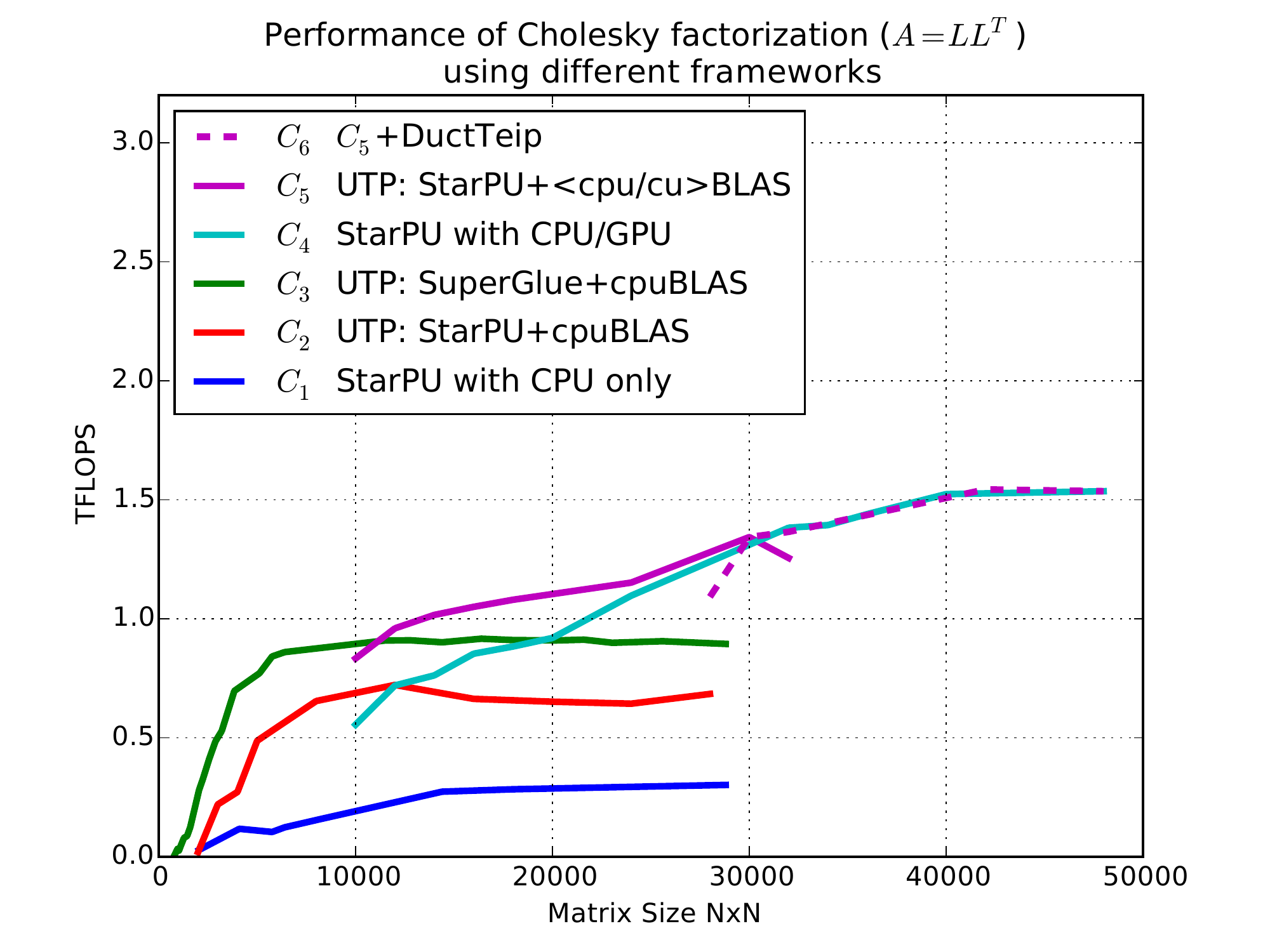}
\label{fig:onenode}
}\quad\subfigure[Relative performance of distributed Cholesky factorization using UTP implementations vs StarPU only.]{
\includegraphics[width=.5\textwidth,height =.25\textheight]{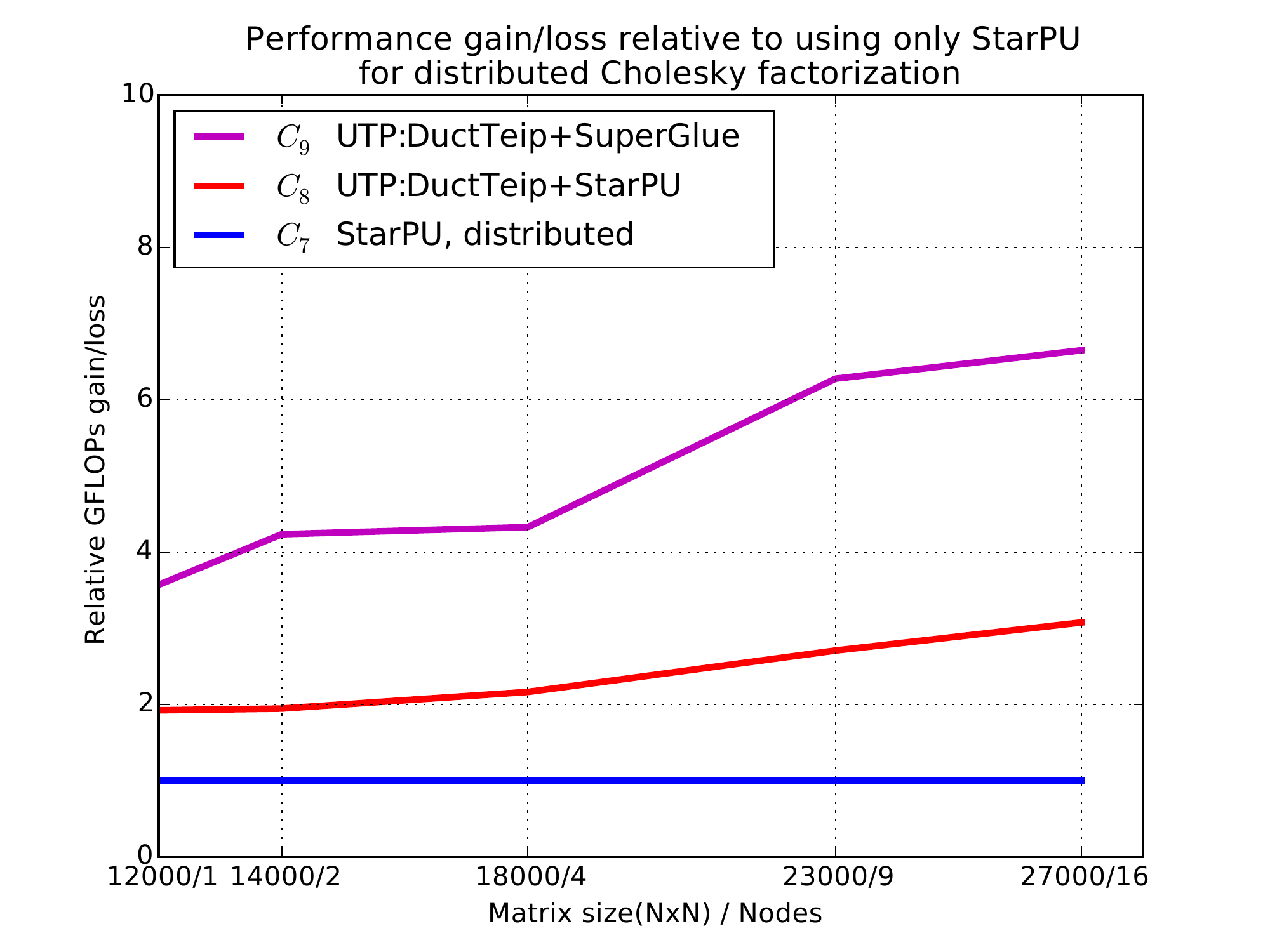}
\label{fig:dist}
}
\caption{Comparing performance of different frameworks with/without the UTP model.}
\label{fig:exp}
\end{figure}

The Cholesky factorization program is executed in multi-core  with or without GPU and multi-node computing environments. The SuperGlue, DuctTeip and StarPU frameworks are used with the UTP and the non-UTP models to compare performance when running the program in these environments, see  Figs.~\ref{fig:onenode} and \ref{fig:dist}. In Fig.~\ref{fig:onenode} different configurations ($C_1$--$C_6$) of frameworks are used for running the Cholesky factorization program in one computing node of multi-cores with/without GPUs. In the configurations  $C_1$--$C_3$ matrices up to 30000$\times$30000 elements are factorized using the StarPU framework ($C_1$), the StarPU wrapper within UTP  ($C_2$) and the SuperGlue wrapper within UTP ($C_3$). For factorizing larger matrices in one computing node, parts of the computations are executed in the GPUs by using the StarPU framework ($C_4$), the StarPU wrapper within UTP ($C_5$) and the DuctTeip wrapper and StarPU wrapper within UTP ($C_6$). The StarPU framework in configuration $C_4$ uses recursive data partitioning and can proceed to larger matrices up to 50000$\times$50000 elements. In configuration $C_5$, the StarPU wrapper is used within UTP without hierarchical data partitioning, and hence cannot factorize larger matrices. However, in configuration $C_6$ adding the DuctTeip wrapper to configuration $C_5$ results in hierarchical data partitioning and enables factorization of larger matrices with the same performance as with the $C_5$ configuration.

In Fig.~\ref{fig:dist}, the performance of different frameworks is compared for Cholesky factorization of matrices in a distributed memory environment. The factorization is performed by the StarPU framework ($C_7$), the DuctTeip and StarPU wrappers within UTP ($C_8$) and the DuctTeip and SuperGlue wrappers within UTP ($C_9$). All the configurations perform hierarchical data partitioning for splitting the factorization into smaller data blocks and as shown in the figure, using frameworks within UTP exhibits better performance than the StarPU only configuration.

These experiments demonstrate that using the UTP model, the single program at the application layer written once and run in several parallel computing environments, not only is independent of any individual framework, but also can attain the most favorable throughput from a customizable mixture of available frameworks.


%


\section{Conclusions and Future Works}\label{sec:conclusion}
We have designed, implemented and verified the UTP model that unifies the cooperation interface between task based frameworks. This interface decouples the frameworks from the program that uses them which enables the program to run in different parallel computing environment, allows independent software development at technical layers and makes the program tolerant to the future changes in the underlying hardware. The configurable task flows in the UTP model allows a program to use a mixture of different frameworks to meet various needs of computations on different computing resources.

\section*{Acknowledgments}
Thanks to Assoc. Prof. Elisabeth Larsson\footnote{http://www.it.uu.se/katalog/bette} for her valuable comments on improving the quality of this paper.
The computations were performed on resources provided by SNIC through
the resources provided by High Performance Computing Center North (HPC2N) under project SNIC2016-7-34.

\bibliographystyle{abbrv}
\bibliography{utp} 

\end{document}